\documentclass{ws-p8-50x6-00}

\begin{document}

\title{D0 Silicon Microstrip Tracker for Run IIa}

\author{E. Kajfasz - \em{For the D0 Collaboration}}

\address{FERMILAB, MS 310, P.O. 500 \\  
Batavia, IL 60510, USA}

\address{CPPM, Case 907, 163 Avenue de Luminy, \\ 
13009 Marseille Cedex 9, FRANCE\\ 
E-mail: kajfasz@fnal.gov}


\maketitle

\abstracts{
We discribe the production, installation and commissioning of the new 
792,576 channel D0 Silicon Microstrip Tracker to be used for the 
2 fb$^{-1}$ of the Run IIa at the Tevatron.}
\section{Introduction}
D0 has built a Silicon Microstrip Tracker (SMT) to help reach its physics 
goals for Run IIa of the Tevatron during which, in the next 2 to 3 years, 
it is supposed to collect 2 fb$^{-1}$ worth of data.\\
Construction of the SMT was finished in December 2000 and installation was
completed for the beginning of Run IIa in March 2001. The following sections
will discuss the different phases of the project.
\section{Design}
\begin{figure}[t]
%\figurebox{20pc}{15pc}{} % to have a box alone
\epsfxsize=25pc % will enlarge or reduce the postscript figures based on the xsize
\begin{center}
\epsfbox{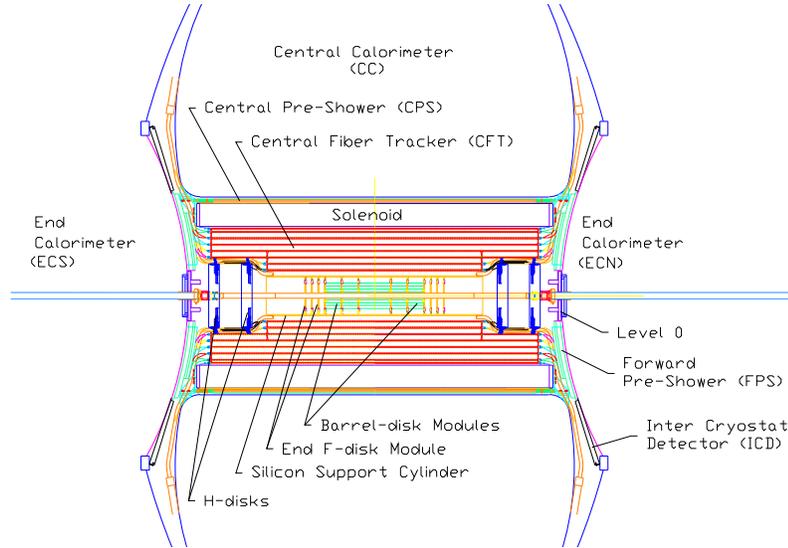} % postscript image file name
\end{center}
\caption{D0 Central Tracking System.  \label{fig:tracking}}
\end{figure}

\begin{figure}
%\figurebox{20pc}{15pc}{} % to have a box alone
\epsfxsize=10pc % will enlarge or reduce the postscript figures based on the xsize
\begin{center}
\epsfbox{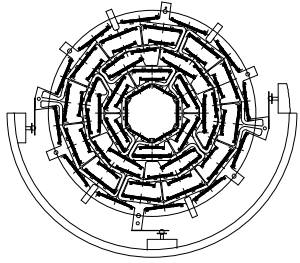} % postscript image file name
\end{center}
\caption{SMT barrel geometry.  \label{fig:barrel}}
\end{figure}

One main improvement included in the D0 detector upgrade~\cite{d0upg} 
for Run IIa is its
central tracking system as shown in Fig.~\ref{fig:tracking}. It includes a 2~T
superconducting solenoid, a Central scintillating Fiber Tracker (CFT) and the
SMT.\\
The SMT design is driven by two classes of events. Barrels and central disks
cover the $\sim 25~cm~RMS$ long luminous region for high $p_T$ central 
physics ($|\eta|<1.5$). Forward disks are implemented mainly to study 
b-physics in the forward region down to
pseudo-rapidities\footnote{Pseudo-rapidity is defined as 
$\eta=-\ln[\tan(\frac{\theta}{2})]$
where $\theta$ is the angle w.r.t. the beam direction.} of 3.
The SMT is comprised of 6 barrels, each barrel mated on one of its ends 
to an F-disk, 2 stacks of 3 F-disks (end disks modules) and 4 H-disks 
(see Fig.~\ref{fig:tracking}). The barrels are 12 cm long and have 72 ladders
arranged in 4 layers (12,12,24,24), each layer having 2 staggered and overlaping sub-layers
(see Fig.~\ref{fig:barrel}). The 2 outer barrels have single sided (SS) and
double sided $2^\circ$~stereo (DS) ladders. The 4 inner barrels have double sided
double metal (DSDM) $90^\circ$~stereo and double sided $2^\circ$~stereo ladders.
The ladders are mounted and aligned to $10-20\mu m$ between 2 precision machined Be bulkheads.
The bulkhead supporting the side of the ladders carrying the read out electronics
is equipped with cooling channels. The F-disks are made of 
12 wedges of double
sided stereo detectors. The H-disks are made of 24 pairs of single sided
detectors glued back to back. For the disks, the wedges are mounted and aligned on Be
rings which include cooling channels. The barrels and F-disks are precisely mounted in 2 carbon 
fiber cylinders which
meet at the nominal interaction point in the D0 detector. The 4 H-disks are
individually mounted in carbon fiber cylinders. Tables ~\ref{tab:radii} and ~\ref{tab:types} 
summarize some SMT design numbers.
\begin{table}[b]
\caption{SMT numbers (module means ladder or wedge).\label{tab:radii}}
\begin{center}
\footnotesize
\begin{tabular}{|l|l|l|l|}
\hline
 &  Barrels& F-disks& H-disks\\
\hline
Channels& 387,072& 258,048& 147,456\\
Modules& 432& 144& 96 pairs\\
Si area& 1.3$m^2$& 0.4$m^2$& 1.3$m^2$\\
Inner radius& 2.7$cm$&2.6$cm$&9.5$cm$\\
Outer radius& 9.4$cm$&10.5$cm$&26$cm$\\
\hline
\end{tabular}
\end{center}
\end{table}
\begin{table}
\caption{SMT detector types (module means ladder or wedge).\label{tab:types}}
\begin{center}
\footnotesize
\begin{tabular}{|l|l|l|l|l|l|l|}
\hline
Location&  Module& Stereo& Pitch& \# of& \#chips& \# of\\
& type& angle ($^\circ$)& ($\mu$$m$)& modules& /mod& HDIs\\
\hline
Barrel layers:& & & & & & \\
L1,L3 (outer bar.) & SS& 0& 50& 72& 3 & 72\\
L1,L3 (inner bar.) & DSDM& 0/90& 50/150& 144& 3/3=6 & 144\\
L2,L4 & DS& 0/$\pm$2& 50/60& 216& 5/4=9 & 216\\
\hline
Disk wedges:& & & & & &\\
F& DS& $+$15/$-$15& 50/60& 144& 8/6 & 288\\
H& SS& $+$7.5/$-$7.5& 50/50& 96& 6/6 & 192\\
\hline
\end{tabular}
\end{center}
\end{table}
Assemblies made of Kapton flex circuits laminated to Be substrates 
(High Density Interconnects or HDIs) are used to hold
the SVXIIe $1.2\mu m$ rad-hard technology read out chips. The SVXIIe has 128 channels, each with
a 32 cell analog pipeline and an 8-bit ADC. It features 53 MHz read out speed, sparsification, 
downloadable
ADC ramp, pedestal, and bandwidth setting~\cite{svxiie}.
\section{Production, assembly, and testing}
HDI flex circuits are electrically tested, laminated to Be substrates, stuffed with component and SVXIIe
chips. The stuffed HDIs are electrically tested for functionality and
performance (pedestal, noise, gain of every channel, sparsification...) and burned in for 2 to 3 days.
In parallel, sensors are tested (CV curves, leakage currents, bias resistors ...) and 
selected using probe stations.
To build a ladder, we use a construction fixture to glue an HDI to silicon sensors. The gluing process is
performed on a CMM to align the sensors within a few microns to the edges of the mounting notches which
reference the ladder w.r.t. the barrel bulkheads. Once the ladders or wedges are glued, their sensors are 
wire bonded
to their SVXIIe read out chips. Altogether, the number of wire bonds in SMT amounts to more than
 1.5 million.
The ladders/wedges are then electrically tested, repaired if necessary, burned in and laser scanned. The
laser scan allows to measure their operating voltage and identify their dead channels. Averaged over all
ladders/wedges, SMT has less than 2\% dead channels.
Selected ladders or wedges are then mounted onto barrel Be bulkheads or disk Be rings with a position
accuracy of about $20\mu m$.
The production started in May 1999 and ended in October 2000. It was mainly 
paced by HDIs fabrication and stuffing problems on one hand, and 
silicon sensor yields, delivery delays and
fabrication problems (e.g. sensor p-stop isolation lithography defects for the DSDM 
sensors, or p-side micro-discharges due to misalignemnt 
of the Al strips w.r.t. the p+ implants, worst in the case of DS $2^\circ$-stereo sensors) 
on the other hand. 
The barrels and disks assembly and their installation in their respective carbon fiber
cylinders were completed by December 2000.
\section{Readout}
\begin{figure}
%\figurebox{20pc}{15pc}{} % to have a box alone
\epsfxsize=10pc % will enlarge or reduce the postscript figures based on the xsize
\begin{center}
\epsfbox{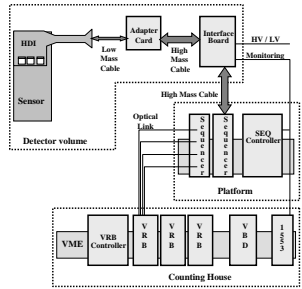} % postscript image file name
\end{center}
\caption{SMT read out.  \label{fig:readout}}
\end{figure}

Fig.~\ref{fig:readout} shows how the read out of the SMT is set up. The HDIs are connected through 2.5m
long Kapton flex cables to Adaptor Cards located on the face of the Central Calorimeter. The ACs transfer the
signals and power supplies of HDIs to 10m long high mass cables which connect to Interface Boards. The
IBs supply and monitor power to the SVXII chips, distribute bias voltage to the sensors and refresh data and
control signals traveling between the HDIs and the Sequencers. The Sequencers control the operation of the
chips and convert their data into optical signals carried over 1Gb/s optical links to VME Readout Buffers
boards. The VRBs receive and hold the data pending a Level-2 trigger decision. The maximum 
L2-accept rate at D0
will be 1KHz, corresponding to a data output rate of $\sim$50Mb/s.
\section{Installation and commissioning}
Barrels and F-disks were installed in the D0 detector by December 2000. The last H-disk was installed early
February 2001. The final cabling was completed in April 2001. Initially, 15\% of the 912 HDIs could not be
read out. However, during the October 2001 Tevatron shutdown, we managed to repair most of them. 95\% 
are now fully functional.
The cooling system was gradually lowered to its nominal temperature to study possible adverse effects on
CFT light yields. The cooling system uses a mixture of 
30\%-glycol/70\%-water circulated at a nominal $-10^\circ$C, so the detectors run between
$-5^\circ$C and $0^\circ$C when powered. 
We optimized the timing and SVXII chip download parameters to maximize the signal to noise ratio.
Calibration procedures and programs (pedestal, noise, gain, sparsification threshold) 
have been implemented. Data was successfully taken with magnet on or off, with the part of the CFT which
was instrumented, and with all the other detector subsystems. Track matching between SMT and CFT shows that
the tracker are inter-aligned within 40$\mu m$.
\section{Conclusions}
The D0 SMT was assembled and installed on time for the start of run IIa. We used
the time until October 2001 to commission and understand the detector hardware and its online/offline
software. Now we are ready to make full usage of it and enjoy the physics goals it will allow us to reach.
By the end of run IIa, after 2fb$^{-1}$, because of radiation damage, the first layers of the SMT will not
be of much use anymore. D0 is already working on a replacement Silicon Microstrip Tracker for Run IIb the
design of wich should allow it to accomodate an integrated luminosity in excess of 15fb$^{-1}$~\cite{ab}.

\end{document}